\documentstyle{EuroPhys}  

\newif\ifboo \boofalse

\input epsf  
  
\newcommand{\hmp}{h^{-1}Mpc}

\newcommand{\be}{\begin{equation}}     
\newcommand{\ee}{\end{equation}}     
\newcommand{\bea}{\begin{eqnarray}}     
\newcommand{\eea}{\end{eqnarray}}     
     
\newcommand{\bef}{\begin{figure}}     
\newcommand{\eef}{\end{figure}}

\def\spose#1{\hbox to 0pt{#1\hss}}      
\def\ltapprox{\mathrel{\spose{\lower 3pt\hbox{$\mathchar"218$}}      
 \raise 2.0pt\hbox{$\mathchar"13C$}}}      
\def\gtapprox{\mathrel{\spose{\lower 3pt\hbox{$\mathchar"218$}}      
 \raise 2.0pt\hbox{$\mathchar"13E$}}}      
\def\inapprox{\mathrel{\spose{\lower 3pt\hbox{$\mathchar"218$}}      
 \raise 2.0pt\hbox{$\mathchar"232$}}}

\begin{document}  
\shorttitle{Gabrielli \& Sylos Labini,  
Fluctuations in galaxy counts}  
  
\title{Fluctuations in galaxy counts:
a new test for homogeneity versus fractality}  
\author{A. Gabrielli \inst{1,3} and F. Sylos Labini\inst{2,3}}  
\institute{  
      \inst{1} Laboratoire de la Physique de la Mati\`ere Condens\'ee,  
\'Ecole Polytechnique, 91128 - Palaiseau Cedex,   France\\  
      \inst{2} D\'ept.~de Physique Th\'eorique,  
            Universit\'e de Gen\`eve 24, Quai E. Ansermet,   
            CH-1211 Gen\`eve, Switzerland \\  
      \inst{3} INFM Sezione di Roma1,  Dipartimento di Fisica, 
      Universit\`a di Roma ``La Sapienza'' -  
            P.le A. Moro 2, I-00185 Roma, Italy\\  
	}  
\rec{ }{ }  
\pacs{  
\Pacs{05}{20$-$y}{Statistical Mechanics}  
\Pacs{98}{65$-$r}{Large scale structure of the Universe}  
      }  
\maketitle

\begin{abstract}  
Fractal properties are usually characterized by    
means of various statistical tools which deal with     
spatial average quantities.    
Here we focus  on the determination of     
fluctuations    
around the average counts and we develop a test  
for the study of galaxy distribution both in      
redshift and  magnitude space.    
Fluctuations in the counts of galaxies, in a  fractal distribution,    
are of the same order of the average number at all scales    
as a function of redshift and magnitude.    
We point out that the study of these kind of fluctuations    
can be a powerful test to understand    
the nature of galaxy clustering at very large scales.    
\end{abstract}

We consider the effect of real space long-range correlations     
in galaxy distribution    
on the statistical properties of counts both in    
redshift  and in magnitude space.    
The existence of large scale structures (LSS)    
and voids in the distribution of galaxies    
up to  several hundreds Megaparsecs is well known for twenty years    
\cite{huchra,tully}. 
The relationship among these structures    
on the statistics of galaxy distribution    
is usually inferred by applying the     
standard statistical analysis as introduced    
and developed by Peebles and coworkers \cite{pee80}.    
Such an analysis {\it assumes} implicitly that     
the distribution is homogeneous at very small scale    
($\lambda_0 \approx 5 \div 10 \hmp$). 
Therefore the system is characterized as having small fluctuations     
about a finite average density.   
If the galaxy distribution had a fractal nature 
the situation would be completely different.     
In this case the average density in finite samples is not 
a well defined quantity: it is strongly sample-dependent 
going to zero in the limit of an infinite volume.      
In such a situation it is not meaningful to study    
fluctuations around the average density extracted from sample data.    
The statistical properties of the distribution should   
then be studied in a completely    
different framework than    
the standard one. We have been working    
on this problem since some time   \cite{slmp98}  
by following the original ideas of Pietronero \cite{pie87}.    
The result is that galaxy structures are    
indeed fractal up to tens of Megaparsecs \cite{joyce99a}.    
Whether   a crossover    
to homogeneity at a certain scale  $\lambda_0$,     
occurs or not (corresponding to the absence    
of voids of typical scale larger than $\lambda_0$)      
is still  a matter of debate  \cite{rees99}.    
At present, the problem is basically that the available redshift surveys    
do not sample scales larger than $50 \div 100 \hmp$    
in a wide portion of the sky and in a complete way.    
In order to clarify this problem  
in the available galaxy catalogs,     
we  propose here to study fluctuations    
around the average behavior of galaxy counts in   
redshift and magnitude space.   
As a matter of fact, in the counts of     
galaxies, large fluctuations around the average behavior 
have been reported by different authors     
both in redshift  \cite{thomas98}  
and photometric surveys     
\cite{picard91,bd97}.      
There have been controversy as to whether they       
are due to real clustering or to     
incompleteness of the catalogs \cite{rees99}   
or to differences       
in the magnitude zero point of the various photometric surveys.    
It is, indeed, possible that discrepancies among these surveys       
are due mostly not to differences in photometric systems       
or in data reduction effects, but rather to real effects,       
i.e. large scale structures. We consider here this  
possibility and        
we present specific calculations about the   
expected value of these fluctuations in three cases:    
(i) a Poissonian homogeneous distribution,   
(ii) a purely fractal distribution, and    
(iii) a homogeneous   
distribution with scale-invariant (i.e. strongly correlated) fluctuations.

Let        
$n(\vec{r})=\sum_i \delta(\vec{r} -\vec{r_i})      
$  
be the microscopic density of a given set of galaxies.      
We {\it assume} hereafter that           
$\nu(L,\vec{r})=\phi(L) n(\vec{r}) $      
is the density of galaxies  in the volume element     
$d^3r$ around the point $\vec{r}$,       
and with luminosity in the range $[L,L+dL]$.      
Note that by writing      
$\nu(L,\vec{r})$ as a product      
of the space density for the luminosity function, the hypothesis     
that galaxy positions are independent     
of galaxy luminosity is implicitly assumed.     
Although it is  known    
that there is a correlation     
between galaxy positions and luminosities   
\cite{binggeli88},  
it has been tested that this is a reasonable     
assumption in the galaxy catalogs available so far     
\cite{slmp98,joyce99a}.     
The galaxy luminosity function usually assumed from observations     
is the so-called Schechter one:        
$\phi(L)= A L^{\delta} \exp(-L/L_*). $   
The parameters     
$L_*$ (luminosity cut-off) and $\delta$ (power law index)     
can be determined experimentally \cite{binggeli88}.    
The constant $A$ is an overall normalizing factor    
such that the integral of $\phi(L)$ over all luminosities    
is equal to one.     
In a volume limited (hereafter VL)
sample \cite{slmp98}    
extracted from a given redshift survey,     
we may compute the number of galaxies as a function of distance of a fixed    
galaxy placed at the origin of coordinates.       
One can write the number-counts in a VL sample     
with faint luminosity limit at $L=L_{VL}$ and   
in a sphere $C(R)$ of radius $R$ in real (redshift) space as    
\be     
\label{e4}     
 N ( < \! R ; \; > \! L_{VL})     
= \int_{L_{VL}}^{\infty}dL\,\phi(L)\int_{C(R)}d^3r\,  n(\vec{r})   \,.   
\ee     
We may also consider the integrated galaxy counts     
as a function of  apparent flux     
(or magnitude), where      
$f=L/4\pi r^2$   
is  the apparent flux.      
In this case one simply have      
\be      
\label{e7}      
N(>f) = \int_f^{\infty} df' n(f') =      
 \int_{0}^{\infty} dL \phi(L) \int d^3r n(\vec{r})       
\Theta\left(\frac{L}{4 \pi r^2}-f \right)  \; ,    
\ee      
where $\Theta(x)$ is the usual step function and 
the spatial integral is extended to the whole space.   
We study below average   
value and fluctuations  of $N(<r)$ (Eq.\ref{e4})     
and $N(>f)$ (Eq.\ref{e7})    
for an homogeneous, a fractal, and an homogeneous distribution    
with scale-invariant fluctuations. 
   
Before going on,   
some statistical definitions about averages and    
correlation functions have to be introduced \cite{pee80}.  
Consider a {\em statistically}   
homogeneous and isotropic particle density    
$n(\vec{r})$ with or without correlations with a well defined average 
value $n_0$. 
Statistical homogeneity and isotropy refer to the fact that  
any $n$-point statistical property of the system  
is a function only on the scalar relative distances between these $n$ points. 
The existence of a well defined average density means that  
\[\lim_{R\rightarrow\infty} 
\frac{1}{\|C(R)\|}\int_{C(R)}d^3r\, n(\vec{r})=n_0>0\]  
(where $\|C(R)\|\equiv 4\pi R^3/3$ is the volume of the sphere  
$C(R)$) independently of the origin of coordinates.  
The scale $\lambda_0$, such that  
$ |\int_{C(R)}d^3r\, n(\vec{r})/\|C(R)\|-n_0|<n_0$ for $R>\lambda_0$, 
is usually called {\em homogeneity scale}. 
If $n(\vec{r})$ is extracted from a density ensemble,  
$n_0$ is considered the same for each realization, i.e. it is  
a self-averaging quantity.   
Let $\left<F \right>$ be the ensemble average of a   
quantity $F$ related to  $n(\vec{r})$. 
If only one realization of $n(\vec{r})$ is available,   
$\left<F \right>$ can be evaluated as an average over   
{\em all} the different points (occupied or not) of the space 
taken as origin   
of the coordinates. 
The quantity   
$\left<n(\vec{r_1})n(\vec{r_2})...n(\vec{r_l})\right>dV_1dV_2...dV_l$   
gives the average probability of finding $l$ particles placed    
in the infinitesimal volumes $dV_1,dV_2,...,dV_l$ respectively   
around $\vec{r_1}, \vec{r_2},...,\vec{r_l}$.   
For this reason   
$\left<n(\vec{r_1})n(\vec{r_2})...n(\vec{r_l})\right>$ is called    
{\em complete} $l$-point correlation function.   
Obviously $\left<n(\vec{r})\right>=n_0$, and in a single sample such that  
$V^{1/3}\gg \lambda_0$, it can be  
estimated  by $n_V=N/V$   
where $N$ is the total number of particle in volume $V$.   
 
Let us analyze the auto-correlation properties of such a system.   
Due to the hypothesis of statistical homogeneity and isotropy,    
$\left<n(\vec{r_1})n(\vec{r_2})\right>$ depends only on   
$r_{12}=|\vec{r_1}-\vec{r_2}|$.   
Moreover,   
$\left<n(\vec{r_1})n(\vec{r_2})n(\vec{r_3})\right>$ is only a function of    
$r_{12}=| \vec{r_1}-\vec{r_2}|$, $r_{23}=| \vec{r_2}-\vec{r_3}|$ and   
$r_{13}=|\vec{r_1}-\vec{r_3}|$.   
The {\em reduced} two-point and three correlation functions $\xi(r)$  
and $\zeta(r_{12},r_{23},r_{13})$ are respectively defined by: 
\bea 
&&\left<n(\vec{r_1})n(\vec{r_2})\right>=n_0^2\left[1+\xi(r_{12})\right]  
\label{03a}\\ 
&&\left<n(\vec{r_1})n(\vec{r_2})n(\vec{r_3})\right>=   
n_0^3  
\left[1+\xi(r_{12})+\xi(r_{23})+  
\xi(r_{13})+\zeta(r_{12},r_{23},r_{13})\right]\,.\nonumber  
\eea 
In order to analyze observations   
from an occupied point it is necessary   
to define another    
kind of average:   
the {\em conditional} average $\left<F \right>_p$.   
This is defined as an ensemble average with the   
condition that the origin of coordinates   
is an occupied point.  
When only one realization of $n(\vec{r})$   
is available,  $\left<F \right>_p$ 
can be evaluated averaging the quantity $F$    
over all the occupied points taken as origin of coordinates.     
$\left<n(\vec{r_1})n(\vec{r_2})...n(\vec{r_l})\right>_pdV_1 dV_2...dV_l$   
is  the average probability of finding $l$ particles placed    
in the infinitesimal volumes $dV_1,dV_2,...,dV_l$   
respectively around $\vec{r_1}, \vec{r_2},   
...,\vec{r_l}$ with the condition that   
the origin of coordinates is an occupied point.   
We call $\left<n(\vec{r_1})n(\vec{r_2})...n(\vec{r_l})\right>_p$   
conditional $l$-point density.   
Applying the rules of conditional probability \cite{feller}, one has:        
\bea 
\label{03c} 
&&\langle n(\vec{r}) \rangle_p  
=\langle n(\vec{0})n(\vec{r})/n_0\\ 
&&\langle n(\vec{r_1})n(\vec{r_2}) \rangle_p =  
\langle n(\vec{0})n(\vec{r_1})n(\vec{r_2})\rangle /n_0\,. 
\nonumber 
\eea 
  
However, in general, the following convention is assumed 
in the definition of the conditional densities:   
the particle at the origin does not observe itself.   
Therefore $\langle n(\vec{r}) \rangle_p$ is defined only for    
$r>0$, and $\langle n(\vec{r_1})n(\vec{r_2}) \rangle_p$  
for $r_1, r_2>0$.    
In the following we use this convention as    
corresponding to the 
experimental data in galaxy catalogs.      
        
     
Imposing a partitioning on the space       
$V$ (in the case $V\rightarrow\infty$) with small cells of volume $dV$,    
let us consider the following Poissonian    
occupation process:      
\be      
\label{p1}      
n(\vec{r})=\left\{ \begin{array}{ll}     
\frac{1}{dV}  & \mbox{with probability}~~~ n_0 dV  \\     
0             & \mbox{with probability}~~~ 1-n_0 dV     
\end{array} \right.       
\ee         
where $dV \ll 1/ n_0$ and $\vec{r}$ is the center of $dV$.   
No correlation is supposed  among different cells.    
First of all one obtains $\langle n(\vec{r}) \rangle = n_0$   
and the homogeneity scale is given by $\lambda_0=1/n_0^{1/3}$.   
The lack of correlations implies    
$\langle n(\vec{r_1}) n(\vec{r_2})  \rangle =   
  n_0^2          \; \;    \mbox{if} \; \;  \vec{r_1} \ne \vec{r_2} $   
and   
$ \langle n(\vec{r_1}) n(\vec{r_2})  \rangle =   
 n_0/dV  \; \; \mbox{if} \; \;  \vec{r_1} = \vec{r_2}   $.  
In the limit $dV \rightarrow 0$ one obtains      
$\xi(r_{12}) = \delta(\vec{r_1}-\vec{r_2})/ n_0  \,,$  
Analogously, one can obtain the three point correlation functions:   
$\zeta(r_{12},r_{23},r_{13})= 
\delta(\vec{r_1}-\vec{r_2})\delta(\vec{r_2}-\vec{r_3})/n_0^2\;.$  
The two previous relations   
say only that there is no correlation    
between different points. That is, the reduced   
correlation functions $\xi$   
and $\zeta$ have only the so called ``diagonal'' part.   
This diagonal part is present in the reduced correlation functions of any    
statistically homogeneous and isotropic distribution with correlations.   
For instance \cite{saslaw} $\xi(r)$ in general can be written as    
$\xi(r)=\delta(\vec{r})/n_0+h(r)\,$,   
where $h(r)$ is the non-diagonal part which is meaningful only for $r>0$.  
In general $h(r)$ is a smooth function of $r$.    
As written before, in the definition of conditional densities,    
we exclude the contribution of the origin of coordinates.   
Consequently, we obtain:    
\bea 
\label{con-poi} 
&&\langle n(\vec{r})\rangle_p=n_0 \\  
&&\langle n(\vec{r_1})n(\vec{r_2})\rangle_p=n_0^2  
\left[1+\delta(\vec{r_1}-\vec{r_2})/n_0\right]\;. 
\nonumber 
\eea 
By using Eq.~(\ref{e4}) and Eq.~(\ref{con-poi}), one can compute both  
the average counts $\langle N(<\!R;\; >\!L_{VL})\rangle_p$  
and the relative fluctuation  
$ \langle \Delta N^2(<\!R;\; >\!L_{VL})  \rangle_p   
  \equiv     
  \langle N^2(<\!R;\; >\!L_{VL})   \rangle_p     
- \langle N(<\!R;\; >\!L_{VL})     \rangle_p^2 \;,$    
obtaining:    
\bea    
\label{p5}    
&&\langle N(<\!R;\; >\!L_{VL})     
\rangle_p = \frac{4 \pi n_0 R^3}{3}     
\int_{L_{VL}}^{\infty} dL\phi(L)\\      
&&\langle \delta_{VL}^2(R) \rangle_p \equiv   
\frac{\langle N^2(<\!R;\; >\!L_{VL})   \rangle_p     
- \langle N(<\!R;\; >\!L_{VL})     \rangle_p^2}    
{\langle N(<\!R;\;>\!L_{VL})\rangle_p^2}     
\sim R^{-3} 
\nonumber 
\eea    
This implies that the typical fluctuations between   
$ N(<\!R;\;>\!L_{VL})$ seen by a single observer    
and the average over all the possible observers   
goes to zero with $R$ as  $R^{-\frac{3}{2}}$.   
In an analogous way, using Eq.~(\ref{e7}) and Eq.~(\ref{con-poi}), 
we can write: 
\bea       
\label{p7}    
&&\langle N(>f) \rangle_p=A n_0 {\cal C} (L_*,\delta)f^{-\frac{3}{2}}\\ 
&&\langle \delta^2(f) \rangle_p \equiv   
\frac{\langle N^2(>f) \rangle_p   - \langle N(>f) \rangle_p ^2  }      
{\langle N(>f) \rangle_p^2  }=    
 \frac{{\cal C}_2 (L_*,\delta)}       
{{\cal C}(L_*,\delta)^2}\frac{1}{n_0}f^{\frac{3}{2}}\,.\nonumber 
\eea 
Then for $f \rightarrow 0$     
the relative fluctuation of the $f$-counts from a single observer and 
the average behavior decreases as $f^{\frac{3}{4}}$.  
The proportionality constants ${\cal C}(L_*,\delta)$ and  
${\cal C}_2(L_*,\delta)$ 
can be evaluated by performing explicitly the integrals 
implied by Eq.~(\ref{e7}):  
\bea 
\label{coeff} 
&&{\cal C}(L_*,\delta) = 1/(3\sqrt{4 \pi}) L_*^{\delta+\frac{5}{2}}      
\Gamma_e\left( \delta + \frac{5}{2} \right) \\ 
&&{\cal C}_2 (L_*,\delta)\! =\! 4 \pi (L_*)^{2 \delta + \frac{7}{2}}       
\int\!\!\int_{0}^{\infty} dx dx' x^{\delta} x'^{\delta}      
e^{-(x+x')}        
\int_{0}^{\infty} dy       
y^2      
\Theta\!\left(\sqrt{\frac{x}{4 \pi }}-y  \right)       
\Theta\!\left(\sqrt{\frac{x'}{4 \pi }}-y  \right)\,, 
\nonumber 
\eea 
where $\Gamma_e(x)$ is Euler gamma-function. 
One can also consider the equivalent result in magnitude space.     
By definition \cite{pee93}        
$f = \frac{L_*}{4 \pi (10 \; pc)^2} 10^{0.4(M_* - m)}     
= K \times 10^{-0.4 m}$.    
Then, from Eq.\ref{p7} one obtains         
\bea 
\label{poi-m} 
&&\langle N(<m) \rangle_p \sim 10^{0.6 m} \\ 
&&\langle \delta^2(m) \rangle_p \equiv 
\frac{\langle N^2(<m) \rangle_p-\langle N(<m) \rangle_p^2} 
{\langle N(<m) \rangle_p^2}\sim 10^{-0.6m}\nonumber 
\eea   
i.e. the normalized fluctuation exhibit an exponential decrease in $m$.


Let us consider the fractal case.   
It is well known that fractal structures      
with the same fractal dimension may have very different      
morphological properties. How such properties can be defined and measured      
in a quantitative way is a very important question.      
Several authors have introduced concepts like lacunarity,      
porosity, log-periodic      
corrections to scaling and others in order to     
go beyond the simple scaling analysis. Here        
we briefly review some useful concepts and definitions.           
For a fractal point distribution with dimension $D<3$    
the conditional one-point density $\left<n(\vec{r})\right>_p$   
(which is hereafter called $\Gamma(r)$) has the following behavior   
\cite{slmp98}     
$\left<n(\vec{r})\right>_p\equiv \Gamma(r)=Br^{D-3}\;. $  
Henceforth the average  mass-length relation (hereafter MLR)      
from an occupied  point is      
\begin{equation}     
\label{f1}     
\langle  N(<R) \rangle_p=     
(4\pi B)/D  \times R^D \; ,     
\end{equation}     
The constant $B$ is directly related to the lower cut-off       
of the distribution.     
Eq.(\ref{f1}) implies that    
the average density in a sphere of radius $R$ around   
an occupied point scales as $1/R^{3-D}$.    
Hence it depends on the sample size $R$, the   
fractal is asymptotically empty and thus    
$\lambda_0\rightarrow\infty$.   
The MLR has a genuine power law behavior   
only when it is averaged over the ensemble, or over all the points     
of the structure if only one realization is available.      
The behavior of  $N(<R)$ from a single point   
presents scale invariant oscillations     
around the average given by Eq.(\ref{f1}).      
The presence of such oscillations is due to the fact that,    
at any scale $R$, the point distribution presents   
voids of radius of the same order as $R$.    
Usually this effect from a single point is described through a     
{\it modulation term} $f(R)$ around the average behavior:      
$N(<R) = (4 \pi B)/D \times  R^D  \times    
 f(R)  \; . $    
$|f(R)|$ is a limited non-decreasing function of $R$.      
In the case of deterministic fractal $f(R)$ can be written as   
\cite{sornette}      
$f(R) = \exp\left(a \; \sin(D_I\log R+\phi)\right) \simeq     
1 + a \sin (D_I  \log R +\phi) \; $, with $a \ll 1$.     
Therefore $f(R)$ is an oscillating function with the wave length    
changing proportionally to the scale $R$.   
In a random fractal, $f(R)$ has a more complex behavior, but keeps these   
qualitative features.   
By averaging over all the observers, $f(R)$ is smoothed out:             
$\langle f(R) \rangle_p = 1  \;.  $      
Note that   
the effect of $f(R)$ is that fluctuations in $N(<R)$ with respect to   
the average   
$\left<N(<R)\right>_p$,    
are proportional to $\left<N(<R)\right>_p$ itself, rather than to its       
square root as in a poissonian distribution.       
Let us now consider the   
two-point conditional density $\langle n(\vec{r_1})n(\vec{r_2})\rangle_p$.   
Blumenthal \& Ball \cite{ball93} have shown    
that statistical translation and rotational invariance,  
 together with scale-invariance, lead to    
\be       
\label{f9}       
\langle n(\vec{r_1})n(\vec{r_2})\rangle_p     
\simeq     
\Gamma(r_1)\Gamma(r_2)       
{\cal L}(r_1/r_2,\theta)   \;,    
\ee     
where   
${\cal L}(r_1/r_2,\theta ) = 1 +       
 g(r_1/r_2, \theta)$, with       
$ \lim_{r_1/r_2 \rightarrow 0 \,, \, r_2/r_1 \rightarrow 0 }    
g(r_1/r_2, \theta) = 0$, is called {\em lacunarity} function.
 
By using  
Eq.~(\ref{e4}), one obtains the average count in a VL sample:    
\bea    
\label{f23}    
&&\langle N(<\!R ;\; >\! L_{VL}) \rangle_p = \frac{4 \pi B}{D}     
 R^D   \times     
\int_{L_{VL}}^{\infty} \phi(L) dL \\  
&&\langle N^2(<\!R ; \; >\!L_{VL})    
\rangle_p - \langle N(<\!R ;\; >\! L_{VL}) \rangle_p^2 
\sim \langle N(<R;>L_{VL}) \rangle_p^2   
\nonumber 
\eea  
on large enough scales, when  shot noise becomes    
negligible.   Then, we expect that fluctuation  
from a single observer in spatial 
counts with respect to the average are of the same order of  
the average itself at any scale. 
Analogously, applying Eq.~(\ref{e7}) to this case, one obtains      
\bea     
\label{f24}      
&&\langle N(>f) \rangle_p  =         
A B {\cal Q}(L_*,\delta) f^{-\frac{D}{2}}\\ 
&&\langle  N^2(> f) \rangle_p -      
\langle  N(> f) \rangle_p^2 =       
A^2B^2  {\cal Q}_2 (L_*,\delta)f^{-D}  \sim \langle  N(> f) \rangle_p^2\,,    
\nonumber    
\eea      
where       
\bea      
\label{f25}      
&&{\cal Q}(L_*,\delta) = (1/(D (4 \pi)^{\frac{D - 2}{2}}))      
L_*^{\delta + \frac{D+2}{2} } \Gamma_e     
\left(\delta + \frac{D+2}{2} \right)\\ 
&& \left\{ 
\begin{array}{ll}   
{\cal Q}_2(L_*,\delta) = (L_*)^{2 \delta + 2 + D}      
\int\!\int_0^{\infty} dx_1dx_2\,     
x_1^{\delta}      
x_2^{\delta}      
e^{- (x_1+x_2)}   
\int_0^{\infty} dy_1 \,  y_1^{D-1}\\   
\times \int_0^{\infty} dy_2\,        
y_2^{D-1}   
\int_{4 \pi} \int_{4 \pi}d\Omega_1 d\Omega_2\,      
g\left( \frac{y_1}{y_2}, \theta \right)       
\Theta\left( \sqrt{\frac{x_1}{4 \pi } }- y_1 \right)       
\Theta\left( \sqrt{\frac{x_2}{4 \pi } }- y_2 \right) 
\end{array} 
\nonumber 
\right. 
\eea     
Eqs.(\ref{f23})-(\ref{f24}) show  
the persistent character of counts fluctuations in the fractal case.    
In terms of apparent magnitudes the normalized fluctuation becomes      
$\langle \delta^2(m)\rangle_p={\cal Q}_2 (L_*,\delta) 
/[{\cal Q} (L_*,\delta)]^2$.  
This implies that counts fluctuations around the average are constant with     
apparent magnitude (or with distance), in contrast with the   
previously considered  poissonian case where    
they are exponentially (or power law) damped.   
A particular attention must be payed to the case $D=3$ in fact, from  
Eqs.(\ref{f23})-(\ref{f24}), while the average counts formulas 
reduce to the Poissonian  
ones, fluctuations stay scale invariant in contrast with  
the Poissonian case. 
This is due to the assumed shape of the lacunarity function,  
which contains in itself the scale-invariant behavior of fluctuations.   
In order to obtain the real  
Poissonian-homogeneous case, it is not enough to consider the limit 
$D\rightarrow 3$, but it is necessary to substitute the relation between 
$2$-point conditional density and $1$-point conditional density 
with $\left<n(\vec{r}_1)n(\vec{r}_2)\right>_p= \left<n(\vec{r}_1)\right>_p 
\left<n(\vec{r}_2)\right>_p$ at large enough $|\vec{r}_1-\vec{r}_2|$, 
and not large enough $r_1/r_2$ which is a weaker condition. 
         

Let us now consider a mixed case in which the system   
is homogeneous (i.e. $\lambda_0$ is finite),   
but presents scale-invariant density fluctuations.   
This last event is in general described by the divergence of   
the  correlation length $r_c$, which is usually   
defined as the scale beyond which $\xi(r)$   
is exponentially damped. It measures up to which   
distance density fluctuations    
density are correlated. Note that while $\lambda_0$   
refers to an one-point property of the system   
(the average density), $r_c$ refers to a two-points   
property (the density-density correlation)   
\cite{perezmercader,gsld00}. Therefore let us consider a system in which   
$\left<n(\vec{r})\right>=n_0>0$   
and $\xi(r)=[\delta(\vec{r})]/n_0+f(r)$, with   
$|f(r)|\sim r^{-\gamma} \;\;\;\;\mbox{for}\;r\gg \lambda_0\,,   
$  
and  $0<\gamma\le 3$.  
The case $\gamma >3$ is not considered here. Indeed, since  
$|\int d^3 r\,\xi(r)|<+\infty$, it is simple to show that fluctuation  
are at maximum Poisson-like, and in the case in which  
$\int d^3 r\,\xi(r)=0$, they can be also smaller.   
Moreover, this case  
corresponds to a scale-invariant fluctuation density field with ``negative''
dimension.   
 
We recall that in the definition of   
$l$-point conditional densities the contribution of the origin   
of coordinates is not considered.   
Therefore $\left<n(\vec{r})\right>_p=n_0[1+f(r)]$.   
In such a situation density fluctuations   
around the average have a scale-invariant nature.   
In fact a density field $\rho(\vec{r})$, defined by   
$\rho(\vec{r})=[n(\vec{r})-n_0]$, represents a kind of fractal    
field whose $1$-point conditional density is   
$n_0\xi(r)\sim r^{-\gamma}$, and $D=3-\gamma$   
represents the fractal dimension.   
For the same reason the quantity    
$\left<\left(n(\vec{r_1})-n_0\right)  
\left(n(\vec{r_2})-n_0\right)\right>_p=n_0^2\left[   
\xi(r_{12})+\zeta(r_1,r_2,r_{12})\right]$  
is analogous to the two-point conditional density of the    
fractal case.   
Therefore one can impose the equivalent of Eq.(\ref{f9}) to the present case:   
$f(r_{12})+\zeta(r_1,r_2,r_{12})=  
f(r_1)f(r_2){\cal L}(r_1/r_2,\theta)$ ,  
where $\theta$ is the angle between $\vec{r_1}$ and $\vec{r_2}$, and    
${\cal L}(r_1/r_2,\theta)$ is   
defined by Eq.(\ref{f9}).   
Therefore, from Eqs.~(\ref{03a}),(\ref{03c}), we can write   
\be   
\left<n(\vec{r_1})n(\vec{r_2})\right>_p-   
\left<n(\vec{r_1})\right>_p\left<n(\vec{r_2})\right>_p =   
n_0\delta(\vec{r_1}-\vec{r_2})+   
n_0^2 f(r_1)f(r_2)g(r_1/r_2,\theta)\,,   
\label{hf3}   
\ee   
where the term in $\delta(\vec{r_1}-\vec{r_2})$ is due to the diagonal part of   
$\xi(r_{12})$. As shown below, this term is important only for   
$\gamma\ge 3/2$ (i.e.    
$D\le 3/2$ if $\gamma>0$).    
In the case of a fractal   
point distribution this contribution was omitted because in    
that case it is always irrelevant.   
At this point we can evaluate counts fluctuations   
around the average both in a spatial (redshift) VL   
sample (for large $R$) and a flux limited one (for small $f$).   
First of all, we obtain (considering only the main contribution):   
$\left<N(<\!R ;\; >\! L_{VL})\right>_p\sim R^3$ and   
$\left<N(>f)\right>_p\sim f^{- \frac{3}{2}} \;.$   
Moreover, if $\gamma<3/2$, we have   
$\langle \delta_{VL}^2(R) \rangle_p \sim R^{-2\gamma}  
\; \;  \mbox{and} \; \; \langle \delta^2(f)\rangle_p \sim f^{\gamma}\,,   
$  
while if $\gamma\ge 3/2$ the contribution from   
$\delta(\vec{r_1}-\vec{r_2})$ in   
Eq.(\ref{hf3}) dominates, and the same result as in the   
Poissonian case is obtained.    
Finally    
in terms of magnitude one find:   
$\langle \delta^2(m) \rangle_p \sim   
10^{-0.4\gamma\,m}\;\;\mbox{ for } \gamma< 3/2$  
and $\langle \delta^2(m) \rangle_p \sim   
10^{-0.6m}\;\;\;\;\mbox{ for } \gamma\ge 3/2$.  
Therefore also in the case of a {\em homogeneous},   
but correlated point distribution,   
normalized counts fluctuations around the average   
must be always damped, no matter the    
large scale behavior of the correlation function.    
We have only a distinction of damping rate between   
the two cases $\gamma\ge 3/2$ (which   
includes also the case of an exponentially tailed $\xi(r)$) and $\gamma <3/2$.   
In the first case, because of the fast decay of   
$\xi(r)$ at large scales, the behavior   
is the same as in the poissonian case without   
correlations; in the second case    
the damping of normalized fluctuations is slower   
as correlations become important.   
We conclude that the only case in which persistent and scale invariant   
normalized counts fluctuations can be observed,    
is the case of a fractal point distribution.    
Note that this behavior can be observed also in the case   
when the point distribution is homogeneous, but we   
are analyzing sample smaller than the    
homogeneity scale $\lambda_0$.     
    
    
We have considered the average and fluctuations of the counts    
in the low redshift approximation. The effect of the cosmological    
corrections (G: geometrical, K: k-corrections, E: luminosity    
evolution correction) are, in the range $0.1 \ltapprox z \ltapprox 1$    
linear with redshift. Such corrections, which are mostly model    
dependent, can change the slope of the average counts determined     
from a single point (i.e. The Earth). For example we have discussed     
the effect of K+G corrections in the determination    
of the slope of the counts as a function of distance    
from one point in \cite{joyce99b}. A similar effect    
can also occur in number counts as    
a function of magnitude in the equivalent regime    
(e.g. $15^m \ltapprox B \ltapprox 20^m$).    
However it is worth to note that the presence of possible persistent and     
scale-invariant fluctuations in the counts    
cannot be due to any smooth correction to     
data as cosmological and evolution effects,   
but they can be the outcome exclusively of     
the strongly correlated effects present in a fractal point distribution.      
{\it In fact,  smooth corrections to the average behavior  
cannot  produce persistent scale-invariant fluctuations in the counts    
of the same order of the average  itself.}    
This property suggests an important experimental   
check of the possible fractal nature   
of the galaxy distribution from the behavior of   
number counts as a function of apparent magnitude  
with the advantage of using also two dimensional   
surveys, rather than only redshift ones.    


{\bf Acknowledgments}   
We thank Y.V. Baryshev, R. Durrer, J.P.Eckmann, P.G. Ferreira, M. Joyce,    
M. Montuori and L. Pietronero for useful discussions.    
This work is partially supported by the      
 EC TMR Network  "Fractal structures and  self-organization"       
\mbox{ERBFMRXCT980183} and by the Swiss NSF.


\end{document}